\documentclass{article}

\usepackage{microtype}
\usepackage{graphicx}
\usepackage{subfigure}
\usepackage{booktabs} 

\usepackage{hyperref}

\usepackage[accepted]{icml2020}

\icmltitlerunning{Cosine Similarity of Multimodal  Content  Vectors for  TV Programmes}

\begin{document}

\twocolumn[
\icmltitle{{Cosine Similarity of Multimodal  Content  Vectors for  TV Programmes}}





\begin{icmlauthorlist}
\icmlauthor{Saba Nazir}{ucl}
\icmlauthor{Taner Cagali}{qmul}
\icmlauthor{Chris Newell}{bbc}
\icmlauthor{Mehrnoosh Sadrzadeh}{ucl}
\end{icmlauthorlist}

\icmlaffiliation{ucl}{Computer Science, University College London, UK, }
\icmlaffiliation{qmul}{Electronic Engineering and Computer Science, Queen Mary University of London, UK, }
\icmlaffiliation{bbc}{Research and Development, British Broadcasting Company, UK}

\icmlcorrespondingauthor{}{saba.nazir.19@ucl.ac.uk}
\icmlcorrespondingauthor{}{m.sadrzadeh@ucl.ac.uk}

\icmlkeywords{Vector Representation, Multimodal Data, Audio, Subscripts, Metadata, Fusion, Recommendation}

\vskip 0.3in
]



\printAffiliationsAndNotice{} 

\begin{abstract}

Multimodal information originates from a variety  of sources: audiovisual files, textual descriptions, and metadata. We show how one can represent the content encoded by each individual  source using vectors, how to combine the vectors via middle and late fusion techniques, and how to compute the semantic similarities between the contents. Our  vectorial representations are built from spectral features and Bags of Audio Words, for audio, LSI topics and Doc2vec embeddings for subtitles,  and the categorical  features, for metadata.  We  implement our model on a dataset of  BBC TV programmes and evaluate the fused representations to provide recommendations. The late fused similarity matrices significantly improve the precision and diversity of recommendations.  
\end{abstract}

\section{Introduction}

Ideas put forwards by Firth and Harris in the 1930's led to the development of vector representations for words. The original word vectors represented  context of textual use  and the cosine distances between them,  semantic similarities \cite{turney2010}. Subsequent research  extended the vector representation methods from words  to sentences and documents; nowadays, these vectors are learnt using neural networks, for a survey see  \cite{JandM}. Recently, the textual vector  representations have  been enriched by other modes of information, such audio-visual and cognitive \cite {bruni2014,kiela2017learning}. The enriched   representations are  also used in multimodal content-based and hybrid recommendation systems for  problems such as cold-start  \cite {oramas2017deep}, \cite {barkan2019cb2cf}, e-commerce assortment  \cite{iqbal2018multimodal} and genre classification \cite{ekenel2013multimodal}. A large body  of work exists here, with none  as extensive or conclusive: e.g.  \cite{Yangetal}, only considers tags and titles for textual data,  \cite{BOUGIATIOTIS} does use full subtitles but does not improve on the metadata-only recommendations. 

This paper works with multimodal vector representations for audio and text, and investigates their application to TV recommendations, based on   a dataset of 145 BBC TV programmes. On the methodological side,  this is the first time that neural (Doc2Vec) and topical (LSI) document vectors are combined with audio (BoAW) vectors and vector representations of metadata. On the experimental side, the late  fused representations significantly improve the precision and diversity of the recommendations. 

\section{Multimodal Content Vectors}

\begin{figure}[t!]
\vskip 0.15in
\begin{center}
\centerline{\includegraphics[width=\columnwidth]{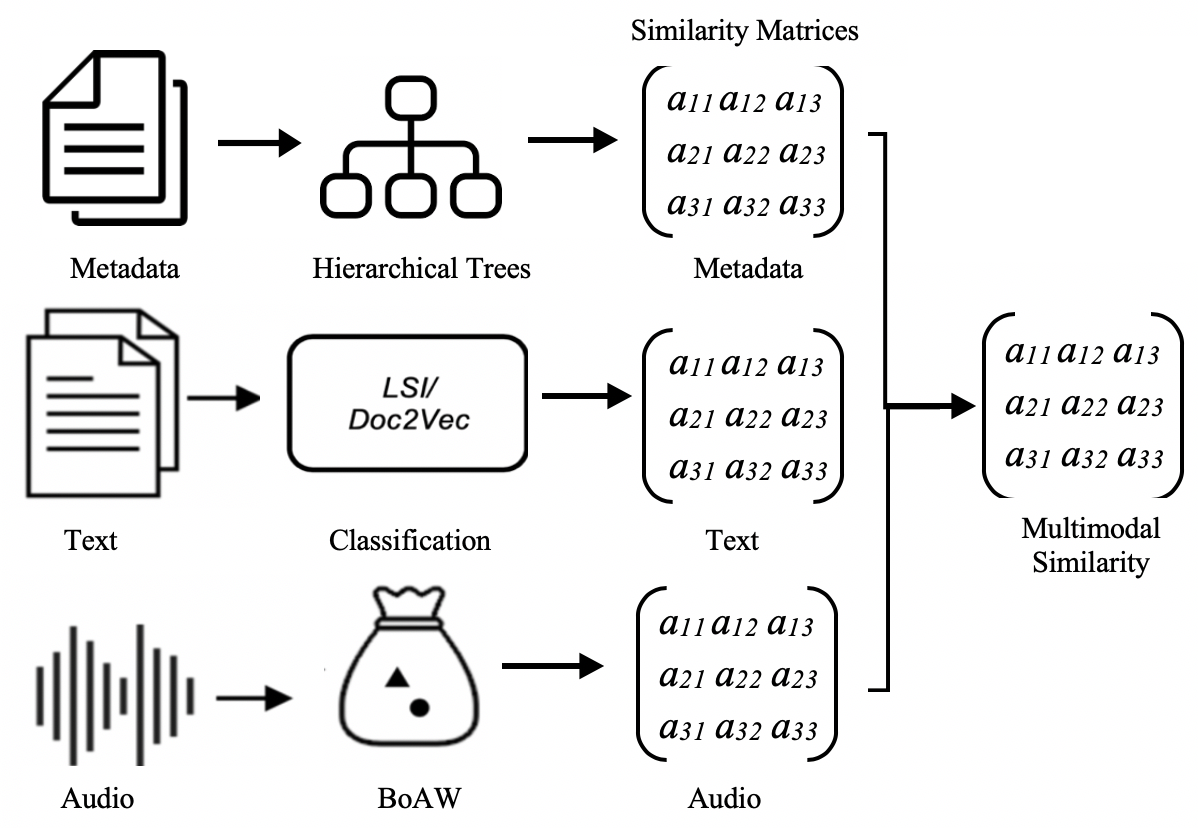}}
\caption{ Multimodal Content Recommendations Framework}
\label{fig: method}
\end{center}
\vskip -0.3in
\end{figure}

Our dataset contains 145 BBC TV programmes with their  subtitle and audiovisual files and metadata information.

{\bf Subtitle Vectors.}   Latent Semantic Indexing (LSI) \cite {papadimitriou2000latent}, a topic modelling technique, was applied to the subtitle files.  LSI is  a two-step procedure. Firstly, a document- term matrix is generated via a low-rank approximation obtained  from the term vector space projections of the Bag of Words vectors. Secondly, Singular Value Decomposition (SVD) is applied to the document-term matrix, where the newly created eigenvectors represent the concepts within the latent space. We worked with  50 dimensional spaces.   LSI improves on the term-document matrices,  but 	does not take word order into account. To deal with this, we worked with neural  semantics embeddings  Doc2vec \cite{le2014distributed}. Doc2vec is an extension of the  neural semantic word embeddings Word2vec \cite{mikolov2013}.    We worked with Paragraph Vector Distributed Memory (PV- DM), which concatenates the unique document ID with the context words with respect to the specified context window over the text  and preserves the order of words.

{\bf  Audio Vectors.} We extracted  acoustic features including MFCCs, Spectral Centroid,  Zero Crossing  Rate, Spectral Flatness and Root Mean Square using LibROSA \cite {mcfee2015librosa}, keeping audio sampling rate of 22050 Hz and hop length of 512 samples, with variable lengths of audio tracks averaging on about 30 mins each for a detailed analysis. The extracted multiple acoustic features are concatenated, normalised and then used as audio vectors for each audio.We then followed  \cite{kiela2017learning}  and  used a  Bag of Audio Words (BoAW)  model  to learn abstract audio vector representations.  BoAW is used in audio information retrieval recognition \cite {liu2010coherent,pancoast2012bag,rawat2013robust} and acoustic event detection \cite {plinge2014bag,grzeszick2015temporal,lim2015robust}. They are learnt via mini-batch K-means clustering with $K=50$.

{\bf  Metadata Vectors.}  Metadata representations  are based on the editorially-assigned attributes of the programmes. We worked with  hierarchical genre information, e.g. "factual/scienceandnature/natureandenvironment", where a match can occur at any level.  A categorical feature vector is created for each attribute by traversing  the trees.

{\bf Fusion.}  Individual content vectors are ranked using cosine similarities and are fused with middle and late  fusion techniques  \cite {kiela2017learning,atrey2010multimodal,zhu2006multimodal}. In middle fusion, we  concatenated the  different vectors representations. In late fusion, we first computed the cosine similarities of pairwise vectors,  resulting in 3 symmetric $145 \times 145$  similarity matrices;  then combined these with each other by weighted averaging. Figure \ref{fig: method} shows our late fusion framework.

\begin{table}[t!]
\caption{Singular and fused model evaluations. The acronyms LSI, D2V, Aud, MD, and Fus are used for Latent Semantic Indexing, Doc2vec, Audios, metadata, respectively. User is the user-based behavioural similarity that we are trying to estimate.}

\label{results-table}
\vskip 0.01in
\begin{center}
\begin{small}
\begin{sc}
\begin{tabular}{lccccr}
\toprule
Model & MAP@10 & ILD@10 & MAP@20 & ILD@20 \\
\midrule
LSI    &11.30&69.89&	13.40	&76.79\\
D2V  & 11.76&77.20&	13.88&	80.37\\
Aud     &06.67  &	77.96  &	8.11	  &81.38\\
MD & 10.78 & 	35.52	 & 12.77 & 	52.72\\
Fus    &{\bf 14.98}	&61.29&	{\bf 17.45}	&70.00\\
\hline

User &15.60 &79.73& 18.51& 80.90\\
\bottomrule
\end{tabular}
\end{sc}
\end{small}
\end{center}
\vskip -0.1in
\end{table}

\section{Evaluation and Results}

Performance of the singular and fused  vectors is evaluated by a personalised Python-based recommender system evaluation framework, developed using MyMediaLite library \cite {MyMediaL39}. We calculated  the Mean Average Precision (MAP) and Intra-list diversity (ILD) of the recommendations at ranks 10 and 20 obtained from cosine similarities,  and compared these with  a metadata-only recommendation system (MD) and the behavioural similarity measure obtained from users’ viewings (USER).  Our aim was  to increase ILD while  maintaining or increasing MAP.  Even individually,  LSI and Doc2Vec  outperformed metadata in both MAP and ILD;  audios (AUD) showed the highest ILD  and lowest MAP. The best performance  was by  the multimodal model late fusion (FUS) with the weights 0.7 LSI,  1.5 D2V,  0.2 AUD, 0.65 MD. It increased  MAP and ILD,  at both ranks 10 and  20, see Table \ref {results-table}. It also came very close to the gold standard user behaviour.

{\bf Analysis.} We analysed individual features of a subset of   programmes to find out  the roles various  features played in measuring the similarities. We worked in three groups: {\bf Group (I).}  closely correlated programmes of the same genre, e.g. Eastenders, Doctors, Waterloo Road. {\bf Group (II).} uncorrelated programmes, of different genres, e.g.   Eastenders, Football League Show, University Challenge. {\bf Group (III).} correlated programmes, of different genres, e.g. Eastenders (soap drama), Jamie Private School Girl (comedy), Notorious Betty Page (autobiographical drama). In {\bf Group (I)}, the genre  similarities were nicely  reflected by the audio and textual features. In {\bf Group (II)} , uncorrelatedness of the programmes were manifested in the textual and audio features. In {\bf Group (III)},   a genre-only analysis failed but a multi modal one succeeded. The similarities between the different-genre programmes  clearly showed themselves in the textual and audio features.   The most efficient differentiating  vectors were BoAW model and   LSI. 

\section{Future Work}
Our framework can easily be extended to other  modalities. We  worked with standard Python packages for sentiment and writing style,   with  format and service metadata,  and with 200 scene images extracted from the video files of  programmes, but did not obtain improvements. Working with more sophisticated  attributes,  and larger number of images  and jointly learning multimodal representations, as in \cite{IqbalKA18},  via neural nets  is  work in progress.

\bibliography{Multimodal-Recommendations}
\bibliographystyle{icml2020}

\end{document}